\documentclass[graybox]{svmult}

\usepackage{mathptmx}       
\usepackage{helvet}         
\usepackage{courier}        
\usepackage{type1cm}        
\usepackage{makeidx}         
\usepackage{graphicx}        
\usepackage{multicol}        
\usepackage[bottom]{footmisc}
\usepackage{amssymb}

\makeindex             

\begin{document}

\title*{An overview of the current status of CMB observations}
\titlerunning{CMB observations}
\author{R.B. Barreiro}
\institute{R.B. Barreiro \at Instituto de F{\'\i}sica de Cantabria
  (CSIC-UC), Avda. de los Castros s/n, 39005 Santander (Spain),
  \email{barreiro@ifca.unican.es}}
%
%
\maketitle

\abstract*{}

\abstract{In this paper we briefly review the current status of 
the Cosmic Microwave Background (CMB)
  observations, summarising the latest results obtained from CMB
  experiments, both in intensity and polarization, and the
  constraints imposed on the cosmological parameters. We also present
  a summary of current and future CMB experiments, with a special
  focus on the quest for the CMB B-mode polarization.}

\section{Introduction}
\label{sec_bb:introduction}

In the last years, a series of high-quality cosmological data sets
have provided a consistent picture of our universe, the so-called
concordance model. This model presents a flat universe with an energy
content of about 70 per cent of dark energy, 25 per cent of cold dark
matter and only around 5 per cent of baryonic matter. The data also
indicate that the primordial density fluctuations are primarily
adiabatic and close to Gaussian distributed with a nearly scale
invariant power spectrum.

The Cosmic Microwave Background (CMB) observations are playing a key role in
this era of precision cosmology (for a recent review see \cite{cha09}). The data collected from a large
number of experiments measuring the intensity and, more recently, the
polarization of the CMB anisotropies are in very good agreement with
the predictions of the inflationary paradigm. Most notably, the NASA
WMAP (Wilkinson Microwave Anisotropy Probe) satellite, launched in
June 2001, has constrained the cosmological parameters down to a
few per cent \cite{kom09}. The detection of the
E-mode polarization of the CMB, first by DASI \cite{kov02} and
later by a handful of experiments, also provided strong support to the concordance model.

The major challenge in current CMB Astronomy is the detection of the primordial
B-mode polarization, which would constitute a direct proof of the
existence of a primordial background of gravitational waves, as
predicted by inflation. A large effort is currently being put within
the CMB community in order to achieve this goal. Some experiments are
already putting limits on the amplitude of the B-mode, while many
others are in preparation. Complementary, a good number of CMB experiments are 
dedicated to the study of the CMB at very small scales, which will provide very valuable
information about secondary anisotropies, such as those due to the Sunyaev-Zeldovich (SZ) effects
or gravitational lensing. Moreover, the ESA Planck satellite
\cite{planck}, that has been launched in May 2009, will provide
all-sky CMB observations, both in intensity and polarization, with
unprecedented sensitivity, resolution and frequency coverage.

Another very active field of research is the study of the temperature
distribution of the CMB. The standard inflationary scenario together
with the cosmological principle predict that the CMB anisotropies
should follow an isotropic Gaussian field. However, alternative
theories predict the presence of non-Gaussian signatures in the
cosmological signal. Interestingly, different works have found
deviations of Gaussianity and/or isotropy in the WMAP data whose
origin, at the moment, is uncertain (see \cite{emg08} for a
review and references therein). Future Planck data is expected to shed
light on the origin of these anomalies.

The outline of the paper is as follows. Section~2 reviews some
recent CMB observational results, both in intensity and
polarization. Section~3 discusses current and future CMB experiments,
including the Planck satellite.

\section{Observational results}
\label{sec_bb:obs}

In the last decade, there has been an explosion of data that has
allowed a strong progress in the characterisation of the CMB fluctuations. 
In particular, the unambiguous detection of the position
of the first peak by different experiments
(Boomerang~\cite{boomerang}, MAXIMA~\cite{maxima}) determined that the
geometry of the universe is close to flat. In subsequent years, other experiments 
such as Archeops \cite{ben03}, VSA \cite{gra03} and, most notably, the
NASA WMAP satellite confirmed these results, imposing strong constraints on the
cosmological parameters. Complementary, other cosmological data sets have also
produced very valuable results, e.g. \cite{fre01,sel06,per07,kow08,rie09}, which 
can be combined with the CMB to produce even tighter constraints \cite{kom09}.
In addition, a series of experiments are measuring the polarization power
spectrum with increasing sensitivity, confirming further the current consistent
picture of the universe.

WMAP consists of five instruments (with a total of 10 
differencing assemblies) observing at frequencies ranging from 23 to 94 GHz, with a
best resolution of 13 arcminutes. The latest published results are
based in 5-year of data, although the satellite continues in
operation.  The WMAP team found that the simple six-parameter
$\Lambda$CDM model -- a flat model dominated by dark energy and dark
matter, seeded by nearly scale-invariant, adiabatic, Gaussian
fluctuations -- continues to provide a good fit to the data. In
addition, the model is also consistent with other cosmological data
sets. Table~\ref{tab_bb:parameters} shows the cosmological parameters
for the simple $\Lambda$CDM model as obtained by~\cite{kom09} using only WMAP and
combining data from WMAP, baryon acoustic oscillations~\cite{per07} and
supernovae \cite{kow08}. Moving beyond this simple model, the combined data
set also constrains additional parameters such as the tensor to scalar
ratio $r < 0.22$ (95 per cent CL) and put simultaneous limits on the
spatial curvature of the universe $-0.0179 < \Omega_k < 0.0081$ and
the dark energy equation of state $-0.14 < 1+w < 0.12$ (both at the 95
per cent CL). It is also interesting to point out that the best current limit on $r$
from CMB data alone is $r < 0.33$ (95 per cent CL) obtained using a combination of WMAP, QUAD and ACBAR
data \cite{bro09}, while the tightest constraint obtained directly from the CMB B-mode
of polarization has recently been provided by BICEP \cite{chi09} and is $r < 0.73$ (95 per cent CL).

\begin{figure}[!t]
\includegraphics[scale=0.64]{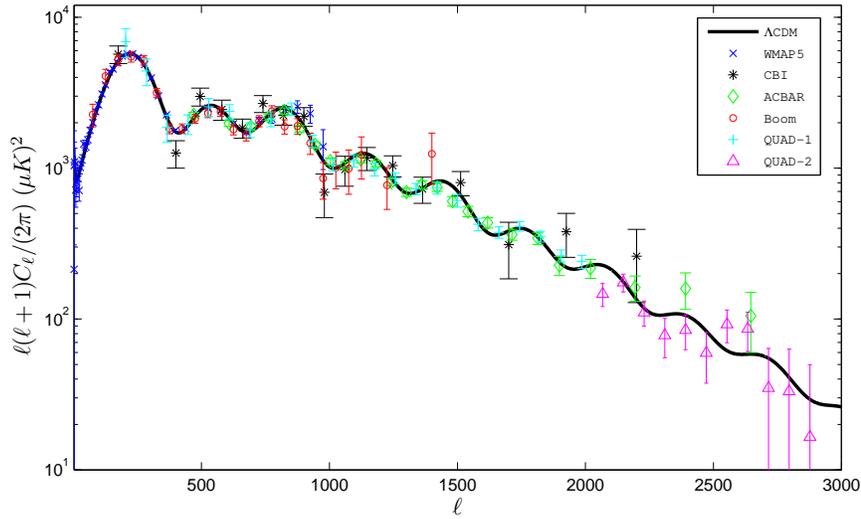}
\caption{CMB temperature power spectrum measured by different experiments: 
WMAP~\cite{nol09}, CBI \cite{sie09}, ACBAR \cite{rei09}, Boomerang \cite{jon06} and QUAD \cite{bro09,fri09}.
The solid line corresponds to the best-fit model obtained using the WMAP 5-yr data \cite{kom09}.}
\label{fig_bb:cl_tt}
\end{figure}

\begin{figure}[b]
\includegraphics[scale=0.64]{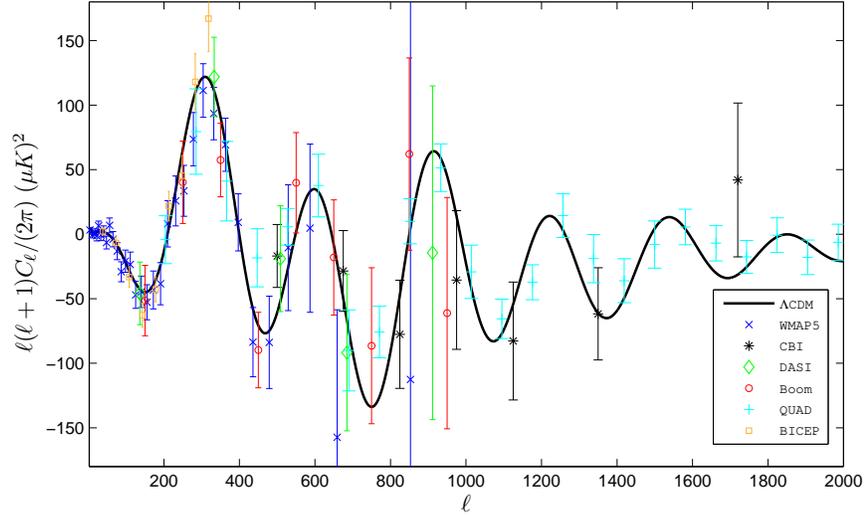}
\caption{TE cross power spectrum measured by different experiments: 
WMAP~\cite{nol09}, CBI \cite{sie07}, DASI \cite{lei05}, Boomerang \cite{pia06}, QUAD \cite{bro09} and BICEP \cite{chi09}.}
\label{fig_bb:cl_te}
\end{figure}

\begin{figure}[b!]
\includegraphics[scale=0.64]{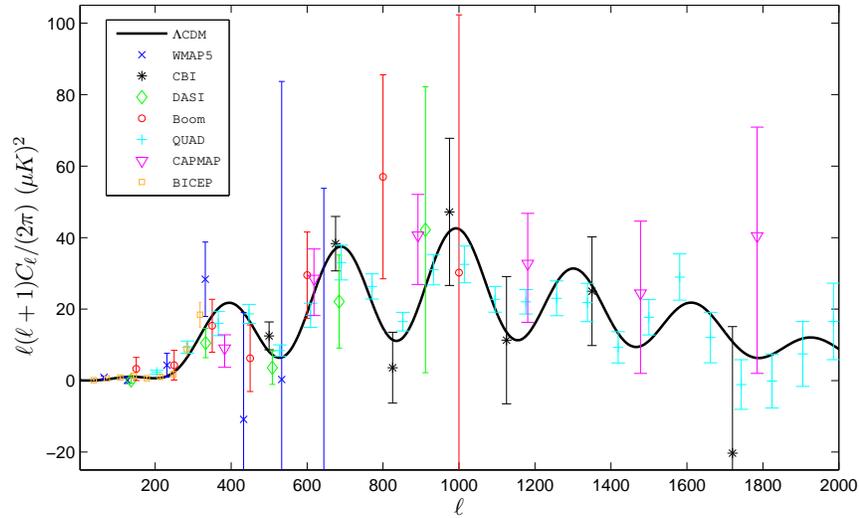}
\caption{CMB E-mode power spectrum measured by WMAP~\cite{nol09}, CBI \cite{sie07}, DASI \cite{lei05}, Boomerang \cite{mon06}, QUAD \cite{bro09}, CAPMAP \cite{bis08} and BICEP \cite{chi09}.}
\label{fig_bb:cl_ee}
\end{figure}

\begin{table}[!t]
\caption{Cosmological parameters, with the corresponding 68 per cent
  intervals, for the 6-parameter $\Lambda$CDM model derived using only WMAP 5-yr data
  and combined WMAP, baryon
  acoustic oscillations and supernovae data (see \cite{kom09} for details). }
\label{tab_bb:parameters}
%
%
\begin{center}
\begin{tabular}{ccc}
\hline\noalign{\smallskip}
Parameter & WMAP & Combined \\
\noalign{\smallskip}\svhline\noalign{\smallskip}
$100\Omega_b h^2$ & 2.273 $\pm$ 0.062 & $2.267^{+0.058}_{-0.059}$\\
$\Omega_c h^2$ & 0.1099 $\pm$ 0.0062 & 0.1131 $\pm$ 0.0034 \\
$\Omega_{\Lambda}$ & 0.742 $\pm$ 0.030 & 0.726 $\pm$ 0.015 \\
$n_s$ & $0.963^{+0.014}_{-0.015}$ & 0.960 $\pm$ 0.013 \\
$\tau$ & 0.087 $\pm$ 0.017 & 0.084 $ \pm $ 0.016 \\
$\Delta_{R}^{2}(k_0)^a$ & $(2.41 \pm 0.11) \times 10^{-9} $ & $ (2.445 \pm 0.096) \times 10^{-9} $ \\
\noalign{\smallskip}\hline\noalign{\smallskip}
\end{tabular}
\end{center}
\hspace{2.4cm}$^a$ $k_0$=0.002 Mpc$^{-1}$.
\end{table}
Fig.~\ref{fig_bb:cl_tt} shows the temperature power spectrum measured
by different experiments.
The solid line is the best-fit $\Lambda$CDM model to the WMAP
5-yr data, which also agrees well with the additional CMB data sets up to
$\ell \approx 2000$. However, some high resolution experiments have found
an excess of power at multipoles $\ell \gtrsim 2000$, in particular,
CBI \cite{sie09} and BIMA \cite{daw06} (which observe at 30
GHz) and, at a lower level, ACBAR \cite{rei09} (at 150 GHz). The
spectrum of the reported excess could be consistent with
Sunyaev-Zeldovich emission from cluster of galaxies but this would
imply a value of $\sigma_8$ larger than the one favoured by
other measurements \cite{kom09,vik09}. Another possible origin of this
excess is the presence of unsubtracted extragalactic sources
\cite{tof05}. Very recently, two experiments, QUAD and SZA, have reported new measurements
of the CMB power spectrum at small scales, finding no excess. In particular, 
QUAD \cite{fri09} reports that, after masking the brightest point sources,
the results at 150 GHz are consistent with the primary fluctuations 
expected for the $\Lambda$CDM model.
The SZA experiment \cite{sha09}, that observes at 30 GHz, finds
that the level of SZ emission is in agreement with the
expected value of $\sigma_8 \approx 0.8$. The latter work also suggests
that the excess found by CBI and BIMA experiments could be due to an
underestimation of the effect of extragalactic point sources.
In any case, further observations will be needed to clarify the origin of
this excess.
Regarding polarization, several experiments have obtained very
valuable data in recent years, providing a further test of the
concordance model. In particular, the large angle anticorrelation seen
by WMAP in the cross power spectrum between temperature and
polarization (TE) implies that the density fluctuations are primarily
adiabatic, ruling out defect models and isocurvature models as the
primary source of fluctuations \cite{pei03}. In addition to WMAP
\cite{nol09}, the TE cross power spectrum has also been
measured by a number of experiments: DASI \cite{lei05}, CBI
\cite{sie07}, BOOMERANG \cite{pia06}, QUAD \cite{bro09} and BICEP \cite{chi09}.
A compilation of these measurements are shown in Fig.~\ref{fig_bb:cl_te}.
Regarding the E-mode of polarization, after its first detection by
DASI \cite{kov02,lei05}, several experiments have delivered further
measurements covering different ranges of angular scales: WMAP
\cite{nol09}, CBI \cite{sie07}, CAPMAP
\cite{bis08}, BOOMERANG \cite{mon06}, QUAD
\cite{bro09} and BICEP \cite{chi09}. Fig.~\ref{fig_bb:cl_ee} shows the E-mode power
spectrum measured by these experiments, where acoustic oscillations
are already seen. Conversely, no detection of the
B-mode polarization has been found up to date, although several
experiments have imposed upper limits, including the polarization
experiments previously mentioned. In particular, BICEP \cite{chi09} (at scales $\gtrsim 1^{\circ}$ ) and QUAD \cite{bro09} 
(at scales $\lesssim 1^{\circ}$) have recently provided the tightest upper limits for the B-mode power
spectrum (for a recent compilation of B-mode constraints see \cite{chi09}).


Although most observational results show consistency with the
concordance model, it is also interesting to point out that QUAD has
recently found some tension between their
polarization data and the simple $\Lambda$CDM model, which seems to be
originated by the TE power spectrum \cite{cas09}. Although this deviation is
not highly significant, it will be interesting to see whether it is
confirmed or not by future polarization experiments.

A number of works have also found deviations from Gaussianity and/or
isotropy in the WMAP data, including, among others, a large cold spot in the southern
hemisphere \cite{vie04,cru07a} north-south asymmetries \cite{eri07,rat07,han08,hof09},
anomalies in the low multipoles \cite{oli04,bie05,chi07,cop07,lan07}, anisotropies in the amplitude
and orientation of CMB features \cite{vie07,wia08}, an anomalously low CMB variance \cite{mon08}
or anomalous properties of CMB spots \cite{lar04,hou09,aya09}. 
Although several possibilities have been considered to explain some of the anomalies, e.g.
\cite{jaf06,cru07b,cru08,ino07,gro09}, their origin is still
uncertain. The future Planck data, with a larger frequency coverage and
better sensitivity than WMAP, as well as a different scanning strategy, will allow one 
to carry out a more detailed study of the temperature distribution of the CMB, helping 
to shed light on these results.

Different groups have also placed constraints on some
physically-motivated non-Gaussian models characterised by the $f_{NL}$
parameter \cite{bar04} finding, in general, consistency with Gaussianity, e.g.
\cite{kom09,cur09,cur09b,pie08,vie09,smi09,hik08}. In particular, the best
limits up to date are $ -4 < f_{NL}^{local} < 80$ \cite{smi09} and $151
< f_{NL}^{equil} < 253$ \cite{kom09}, for the local and equilateral
models respectively, at the 95 per cent CL. However, \cite{yad08} have
found a deviation from the Gaussian hypothesis at the 2.8$\sigma$ for the local
model, in disagreement with the previous mentioned results. Planck
data, as well as future WMAP data with higher sensitivity, will help to
confirm or discard the presence of such deviation.

It is also interesting to point out that the CMB polarization, and in particular the TB and EB 
cross-correlation spectra, can also be used to search for possible signatures of parity violation, e.g. \cite{kom09,xia08}.

\section{Summary of CMB experiments}
\label{sec_bb:expt}

The most notable CMB experiment to operate in the near future is the
ESA Planck satellite \cite{planck}, that has been launched in May 2009.
Planck will measure the CMB fluctuations over the whole sky, in
intensity and polarization, with an unprecedented combination of
sensitivity ($\Delta T/T \sim 2 \times 10^{-6}$), angular resolution
(up to 5 arcminutes), and frequency coverage (30-857 GHz). The main
characteristics of Planck are summarised in
Table~\ref{tab_bb:planck}. Planck will allow the fundamental
cosmological parameters to be determined with a precision of $\sim 1$
per cent and will set constraints on fundamental physics at energies
larger than $10^{15}$ GeV, which cannot be reached by any conceivable
experiment on Earth. In addition, it will provide a catalogue of
thousands of galaxy clusters through the SZ effect
and very valuable information on the properties of radio and infrared
extragalactic sources as well as on our own galaxy.

\begin{table}

\caption{Summary of Planck instrument characteristics (taken from
  \cite{planck}) }
\label{tab_bb:planck}
%
\begin{center}
\begin{tabular}{|l|ccc|cccccc|}
\hline  
& \multicolumn{3}{c}{LFI} & \multicolumn{6}{|c|}{HFI} \\ 
\hline 
Detector Technology & \multicolumn{3}{c}{HEMT arrays} & \multicolumn{6}{|c|}{Bolometer arrays} \\ 
Center Frequency (GHz) & 30 & 44 & 70 & 100 & 143 & 217 & 353 & 545 & 857 \\ 
Angular Resolution (arcmin) & 33 & 24 & 14 & 10 & 7.1 & 5.0 & 5.0 & 5.0 & 5.0 \\ 
$\Delta T/T$ per pixel (Stokes I)$^a$ & 2.0 & 2.7 & 4.7 & 2.5 & 2.2 & 4.8 & 14.7 & 147 & 6700 \\ 
$\Delta T/T$ per pixel (Stokes Q $\&$ U)$^a$ & 2.8 & 3.9 & 6.7 & 4.0 & 4.2 & 9.8 & 29.8 & -- & -- \\ 
\hline 
\end{tabular} 
\end{center}

$^a$ Goal (in $\mu$K/K) for 14 months integration, 1$\sigma$, for square pixels whose sides are given in the row angular resolution.
\end{table}

Complementary, a good number of ground-based and balloon-borne
experiments are operating, or in preparation, in order to measure the
intensity and polarization of the CMB with increasing sensitivity and
resolution. Some of these experiments are devoted to the study of the
CMB fluctuations at very small scales (a few arcminutes or below) and, in particular,
to the study of the CMB secondary anisotropies, including those produced by the 
SZ effects and gravitational lensing. This will allow a further test of the 
concordance model as well as to clarify the possible excess of power found at 
small angular scales by previous CMB observations. Within this type of experiments we can mention 
AMI \cite {zwa08}, SPT \cite{kos06}, ACT \cite{ruh04} or AMiBA \cite{wu09}.

\begin{table}[b]
\caption{Summary of the main characteristics of some B-mode polarization experiments}
\label{tab_bb:pol}
\begin{center}
\begin{tabular}{|l|c|c|c|c|}
\hline
& Angular resolution & Frequency & Goal & Starting\\ 
~~Experiment & (arcmin) & (GHz) & ($r$) & Year \\
\hline
\multicolumn{5}{|c|}{Ground Based} \\
\hline
ABS \cite{sta08} & 30 & 145 & 0.1 & 2010 \\
BRAIN \cite{cha08} & $\sim 60$ & 90, 150, 220 & 0.01 & 2010  \\ 
C-BASS \cite{pea07} & 51 & 5 & -- & 2009 \\
Keck Array \cite{ngu08} & 60 -30 & 100, 150, 220 & 0.01 & 2010 \\
MBI \cite{tuc08}  &  $\sim 60$ & 90 & -- & 2008 \\ 
QUIET \cite{sam08} &  28 - 12 & 40, 90 & 0.01 & 2008 \\ 
QUIJOTE \cite{rub08} &  55 - 22 & 11, 13, 17, 19, 30 & 0.05 & 2009 \\ 
PolarBear \cite{lee08} & 4 - 2.7 & 150, 220 & 0.025 &  2009 \\ 
\hline
\multicolumn{5}{|c|}{Balloon Borne} \\
\hline
EBEX \cite{gra08} & 8 & 150, 250, 410 & 0.02 & 2009 \\ 
PAPPA \cite{kog06} &  30 & 90, 210, 300 & 0.01 & 2010 \\ 
PIPER &  $\sim 15$ & 200, 270, 350, 600 & 0.007 & 2013 \\
SPIDER \cite{cri08} & 58 - 21 & 100, 145, 225, 275 & 0.01 & 2010 \\ 
\hline
\multicolumn{5}{|c|}{Satellite} \\
\hline
Planck \cite{planck} & 33 - 5 & 30 - 353 & 0.05 & 2009   \\ 
\hline
\end{tabular} 
\end{center}
\end{table}
However, the major challenge of current CMB Astronomy is
the detection of the primordial B-mode polarization, which will
imply the existence of a primordial background of gravitational waves,
as predicted by inflation. Table~\ref{tab_bb:pol} summarises some of the main on-going and future
experiments targeted to study the CMB B-mode polarization. For comparison, we also include
the Planck satellite 
in the table, as well as the C-Bass experiment which is devoted to the study of 
the synchrotron polarization and will provide complementary information to 
other experiments. The different experiments cover a wide range of frequencies, resolutions and 
technologies and will allow to detect (or to constrain) values of $r \approx 0.01$ in the next few years.
In addition, design studies for the next generation of satellite missions are being conducted 
(BPol \cite{ber09}, EPIC\cite{boc08}), 
which aim to achieve a sensitivity of $r \approx 0.001$, provided that foreground contamination can
be properly removed.

\section{Conclusions}
During the last years, a consistent picture of our universe, the
so-called concordance model, has emerged due to the availability of
several high quality data sets. In particular, CMB observations have
significantly contributed to improve our description of the
universe. However, some fundamental questions still remain to be
answered such as which is the nature of dark matter and dark energy,
which parameters characterise the inflationary era or which is the
origin of the WMAP anomalies. The future CMB data from the Planck
satellite, as well as from other CMB experiments, will help to answer
these open questions. In addition, the quest for the B-mode of
polarization has already started and, if the scalar-to-tensor ratio is
$r \approx 0.01$ or larger, the primordial background of gravitational waves --
expected from inflation -- could be detected in the next years. This
would constitute a major breakthrough in our understanding of the early
universe.

\begin{acknowledgement}
The author thanks Patricio Vielva and Enrique Mart{\'\i}nez-Gonz\'alez for a careful reading of the manuscript.
I acknowledge partial financial support from the Spanish Ministerio de Ciencia e Innovaci\'on 
project AYA2007-68058-C03-02.

\end{acknowledgement}
\bibliographystyle{spmpsci}
\bibliography{refs_bb}

\end{document}